\documentclass[a4paper,11pt]{article}
\usepackage[utf8]{inputenc}
\usepackage[linktocpage,unicode]{hyperref}
\usepackage{csquotes,graphicx}
\usepackage[export]{adjustbox}
\usepackage[linktocpage,unicode]{hyperref}
\usepackage{url}
\usepackage{indentfirst}
\usepackage{xcolor}
\usepackage{amssymb,amsmath}
\usepackage{geometry}
\usepackage{subcaption}
\usepackage{cite}

\def\nqq{\hspace*{-2em}}

\def\lal{&&\nqq {}}
\def\eq{Eq.\,}
\def\eqs{Eqs.\,}
\def\beq{\begin{equation}}
\def\eeq{\end{equation}}
\def\bear{\begin{eqnarray}}
\def\bearr{\begin{eqnarray} \lal}
\def\ear{\end{eqnarray}}
\def\earn{\nonumber \end{eqnarray}}

\def\nnn{\nonumber\\ \lal }
\def\nnnv{\nonumber\\[5pt] \lal }

\def\yyy{\\[5pt] \lal }

\def\mn{_{\mu\nu}}
\def\MN{^{\mu\nu}}

\def\e{{\,\rm e}}
\def\d{\partial}

\def\const{{\rm const}}
\def\then{\ \Rightarrow \ }

\def\Half{\dfrac 12}
\def\half{\tfrac 12}

\def\S{{\mathbb S}}

\def\M{{\mathbb M}}
\def\oR{{\overline R}}
\def\od{{\overline \d}}
\def\og{{\overline g}}
\def\cV{{\cal V}}

\def\sss{\scriptscriptstyle}
\def\mD{m_{\sss D}}

\usepackage{color}

\def\rf{\eqref}
\def\eqn{\eq\eqref}

\title{Local regions with expanding extra dimensions}

\small
\author{ {Kirill A. Bronnikov} \\
\it	VNIIMS, Ozyornaya ul. 46, Moscow, 119361, Russia; \\
\it	Institute of Gravitation and Cosmology, RUDN University,\\ 
\it	ul. Miklukho-Maklaya 6, Moscow, 117198, Russia; \\
\it	National Research Nuclear University ``MEPhI'',\\
	Kashirskoe shosse 31, Moscow, 115409, Russia\\[10pt]
{Sergey G. Rubin} \\
\it	National Research Nuclear University ``MEPhI'', \\ 
\it	N.I. Lobachevsky Institute of Mathematics and Mechanics,\\
\it	Kazan  Federal  University}
\date{}

\begin{document}
\maketitle
\normalsize

\begin{abstract}
	We study possible spatial domains containing expanding extra dimensions. We show that 
	they are predicted in the framework of $f(R)$ gravity and could appear due to quantum 
	fluctuations during inflation. Their interior is characterized by the multidimensional curvature 
	ultimately tending to zero and a slowly growing size of the extra dimensions.
\end{abstract}

\section{Introduction}

  It is usually assumed that the nucleation of our Universe is related to quantum processes at the 
  Planck energy scale \cite{1994CQGra..11.2483V,Bousso:1998ed,2009arXiv0909.2566H}.
  After nucleation, various manifolds evolve classically, forming a set of manifolds, one of which 
  is our Universe. It is a serious problem to fix the Lagrangian parameters to satisfy the 
  observations. The complexity of this problem is aggravated by inclusion of extra 
  dimensions. The latter are of particular interest because the idea of extra space is widely used 
  in modern research dealing with such problems as grand unification 
  \cite{ArkaniHamed:1998rs,1999NuPhB.537...47D}, the cosmological constant problem 
  \cite{Sahni:1999gb,2003PhRvD..68d4010G,Rubin:2016ude} and so on. 
  The assumption of extra space compactness immediately leads to the question: why is a
  specific number of dimensions stable or slowly expanding 
  \cite{2007JHEP...11..096G, 2002PhRvD..66b4036C, 2002PhRvD..66d5029N}.
  Stabilizing factors could be, for example, scalar fields 
  \cite{2017PhRvD..95j3507K,2007JHEP...11..096G} 
  or gauge fields \cite{2009PhRvD..80f6004K}. A static solution can be obtained using the 
  Casimir effect \cite{1984NuPhB.237..397C,2018GrCo...24..154B} or form fields 
  \cite{1980PhLB...97..233F}. 
  A contradiction with observations can be avoided if the extra-dimensional scale factor $b(t)$ varies 
  sufficiently slowly \cite{1984PhRvD..30..344Y,2013GReGr..45.2509B}.

  One of the ways to stabilize the extra dimensions is based on gravity with higher derivatives, 
  which is widely used in modern research. One of the most promising models of inflation is 
  the Starobinsky model using a purely gravitational action \cite{qua1}. 
  Attempts to avoid the Ostrogradsky instabilities are made \cite{2017PhRvD..96d4035P}, 
  and extensions of the Einstein-Hilbert action attract much attention.  Our model contains 
  $f(R)$ gravity with addition of the Kretchmann invariant and the Ricci tensor squared. Its 
  action can be considered as a basis for an effective field theory 
  \cite{Donoghue:1994dn,2007ARNPS..57..329B}

  The recent paper \cite{Lyakhova:2018zsr} studied the evolution of manifolds after their \
  creation on the basis of a pure gravitational Lagrangian with higher derivatives. The final metrics 
  may differ in different spatial regions if the model admits several stationary states. It is 
  precisely the case for the model discussed in \cite{Fabris:2019ecx} where one of the 
  stationary states was studied. The Lagrangian parameters at low energies are chosen in 
  such a way as to supply an (almost) Minkowski space for the present Universe, the stationarity 
  of the extra-space metric, and to reproduce an inflationary stage of the expansion with the 
  Hubble parameter of the order of $H\sim 10^{13}$ GeV. Here we will
  discuss another final state admitted by our model, with a comparatively small curvature of the 
  extra dimensions which can still be compatible with observations. Some 3D spatial regions in the 
  Universe could be characterized by such a metric, and it is of interest to study this possibility.

\section{Outlook}

\begin{figure}[t]
\centering
\includegraphics[width=8cm]{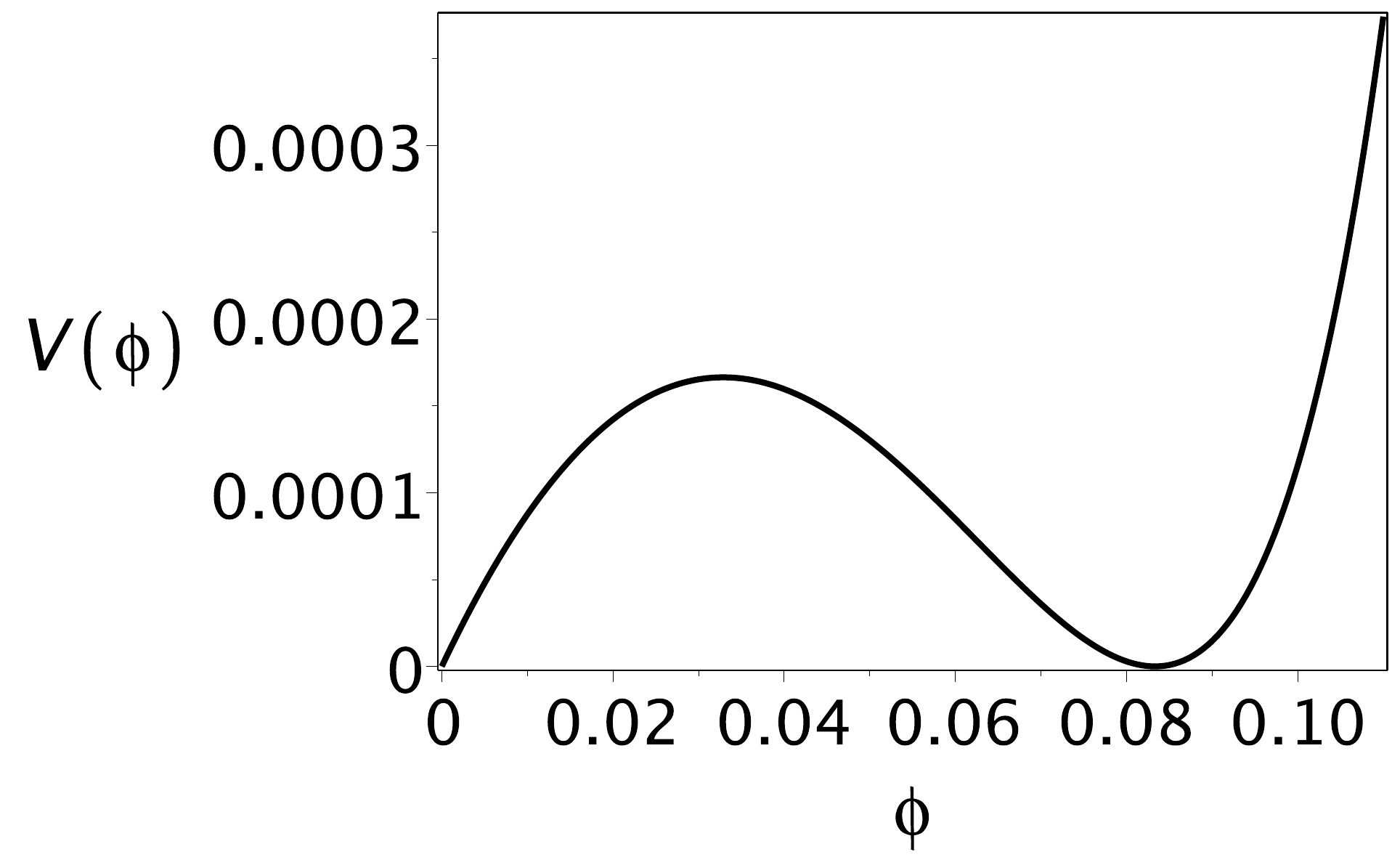} 
\caption{\small
	The effective potential for a viable version of the model \rf{act1}.
	The minimum of the potential is at the point $\phi_m  \simeq  0.083$, 
	$m_D=1, m_D\sim 0.1 M_{\rm Pl}$
	}\label{VK}	
\end{figure}

  Self-stabilization of the extra dimensions is one of necessary elements of models based on 
  compact extra spaces. It has been shown in \cite{BR-06,BR-book}
  that the models with higher derivatives could lead to stationary solutions. The analysis was based on 
  the model where the initial action was taken in the form \cite{BB16, BB17}
\bear                                                         \label{act1}
	S = \Half \mD^{D-2} \int\sqrt{g_D}\,d^{D}x\,
	\big[f(R) + c_1 R^{AB}R_{AB} + c_2 R^{ABCD}R_{ABCD} + L_m\big],
\ear  
  where capital Latin indices cover all $D$ coordinates, $g_D = |\det(g_{MN})|$,
  $F(R)$ is a smooth function of the $D$-dimensional scalar curvature $R$, 
  $c_1,\ c_2$ are constants, $L_m$ is a matter Lagrangian, and $\mD= 1/r_0$ is the 
  $D$-dimensional Planck mass, so that $r_0$ is a fundamental length in this theory.\footnote
  		{Throughout this paper we use the conventions on the curvature tensor 
  		$R_{ABC}^D=\d_C\Gamma_{AB}^D-\d_B\Gamma_{AC}^D
  		+\Gamma_{EC}^D\Gamma_{BA}^E-\Gamma_{EB}^D\Gamma_{AC}^E$ 
  		and for the Ricci tensor $R_{MN}=R^F_{MFN}$. The metric signature $(+ - - \cdots)$
  		and the units $\hbar = c = 1$ are also used.}
  As $L_m$, we can consider the Casimir energy density in space-time with the metric
\beq                                                                \label{ds}
	ds^2 = g\mn dx^\mu dx^\nu - r_0^2 \e^{2\beta(x^\mu)} d\Omega_n^2
\eeq  
  where $x^{\mu}$ are the observable four space-time coordinates, and
  $d\Omega_n^2$ is the metric on a unit sphere $\S^n$. The space-time is a direct 
  product $\M_4\times \M_n$. The function $f(R)$ is taken in a general quadratic form,
\beq       \label{f-R}
	f(R) = a_2 R^2+R - 2\Lambda_D.
\eeq
  Assuming $L_m=0$, we have obtained an effective scalar-tensor theory with the potential 
  presented in Fig.\,\ref{VK}. Here, the internal Ricci scalar $R_n$ actually plays the role of 
  a scalar field $\phi$. Details can be found in  \cite{BR-06,Fabris:2019ecx}

  According to the experimental data, the scale of extra dimensions cannot be larger than 
  $10^{-17}-10^{-18}$ cm. For example, the Ricci scalar $R_n=\phi\sim 0.01 M_{\rm Pl}^2$
  from Fig.\,\ref{VK} is very large, so that the inequality 
\beq   						\label{ineq1}
			R_n\gg R_4
\eeq
  looks natural. Here $R_n$ and $R_4$ are the Ricci scalars of the extra dimensions and of our 
  4D space-time, respectively. The inequality \eqref{ineq1} was used in the derivation of the effective 
  potential in Fig.\,\ref{VK}.

  The scalar field could tend to any minimum, depending on the initial conditions. Hence, different
  spatial regions could be filled with different states of $\phi$, that with $R_n=\phi_m$ or 
  with a small Ricci scalar $R\to 0$ due to the metric fluctuations at high energies.
  
  The metric evolution around the solution $R_n=\phi_m$ could reproduce the inflationary scenario. 
  One can adjust the model parameters to describe the observational properties of the inflationary 
  stage. As usual, strong fine tuning is necessary to produce a small, almost zero 4D cosmological 
  constant.

  In this paper, we intend to analyze the space-time metric that can emerge near the minimum 
  $R\to 0$. Naively, an effective 4D cosmological constant should tend to zero automatically in such 
  regions of space. 
  Is it applicable for life formation? The question is important even if the answer is negative. Even if 
  we live at the standard minimum where  $R_n=\phi_m$, some neighboring regions could be 
  filled with another minimum where $R\to 0$.

  Notice that we are interested in those regions where the Ricci scalar tends to zero. 
  In this case, the inequality \eqref{ineq1} may look suspicious, but it should evidently hold if
  the extra dimensions are spherical and small enough to be invisible by modern instruments.
  But more than that, close to the limit $R\to 0$ one more inequality,
\beq 				\label{Rsmall}
				a_2 R^2 \ll R,
\eeq
  must hold, which strongly facilitates the analysis.

\section{Field equations}
  
  Under the assumption \rf{Rsmall} we can simply put $F(R) = R - 2\Lambda_D$ and neglect
  all the curvature-nonlinear terms in \rf{act1}.
   
  The action \rf{act1} is then reduced to 4 dimensions, in which it takes the form inherent 
  to a scalar-tensor theory specified in 4D space-time with the metric $g\mn$:
\beq  		\label{act2}
		S = \Half \cV \mD^2 \int \sqrt{g_4}d^4 x \e^{n\beta} 
				\Big[R_4 + \frac{n(n-1)}{r_0^2} \e^{-2\beta} + 2n \Box\beta
				+ n(n+1) (\d \beta)^2 - 2\Lambda_D \Big],    
\eeq  
  where $(\d \beta)^2 = g\MN \beta_{,\mu} \beta_{,\nu}$, $\Box = \nabla_\mu \nabla^\mu$, and 
  $\cV(n) = 2\pi^{(n+1)/2}\big/\Gamma(\half(n+1))$  is the volume of a unit sphere $\S^n$.
    
  The standard transition to the Einstein frame with the 4D metric 
\beq
		\og\mn = \e^{n\beta} g\mn	
\eeq  
  brings the action to the form (up to a full divergence) 
\beq  		\label{act3}  
           S = \Half \cV(n) \mD^2 \int \sqrt{\og}\,d^4x\, 
              \Big[\oR_4 + \Half n(n+2) (\od \beta)^2 + \frac{n(n-1)}{r_0^2} \e^{-(n+2)\beta}
              		-  2 \e^{-n\beta} \Lambda_D\Big] + S_m,
\eeq  
   where overbars mark quantities obtained from or with $\og\mn$. 
  
   Now, if we use \rf{act3} to describe cosmological models with the Einstein-frame metric
\beq                 \label{ds_E}
		ds_E^2 = dt^2 - a^2(t) d{\vec x}^2,
\eeq  
   and the effective scalar field $\beta(t)$, we have the Einstein-scalar equations 
   (only two of them are independent, and $V_{\beta} = dV/d\beta$):
\bearr    \label{ddb}
    \ddot\beta + 3\frac{\dot{a}}{a}\dot\beta +\frac 1{ n(n+2)}V_{\beta} = 0, 
\yyy	   \label{E00}	
	3 \frac{{\dot a}^2}{a^2} = \Half n(n+2) {\dot\beta}^2 + V(\beta),
\yyy		   \label{E-tr}
		\frac{6}{a^2} \Big({\dot a}^2 + a\ddot a\Big) =  -n(n+2) {\dot\beta}^2 - 4 V(\beta),
\ear  
  with the potential $V(\beta)$ given by
\bearr  		\label{V-beta}
  	     r_0^2 V(\beta) = \lambda \e^{-n\beta} - k_1 \e^{-(n+2)\beta} + k_C \e^{-(2n + 4)\beta},
\nnn  		
  			\lambda = r_0^2 \Lambda_D, \qquad k_1 = \Half n (n-1).
\ear  
  The last term is related to the Casimir effect, with $k_C$ of the order $10^{-3} - 10^{-4}$,
  see \cite{BB16, BB17} and references therein. 
  
  As is shown in the next section, the first term in the potential \eqref{V-beta} dominates.

\section{Can we live in a region with $R\to 0$? Some estimates}
  
  Consider the dimensionless version $W(x)$ of the potential $V(\beta)$, with $x = \e^{-\beta}$,
\beq              \label{W-x}
  		W(x) = \lambda x^n - k_1 x^{n+2} + k_C x^{2n+4}.
\eeq  
    
  Let us determine the order of magnitude of the variable $x$ such that the condition \rf{Rsmall}
  will be satisfied. It can be estimated by analogy with the well-known 4D models that employ
  $f(R)$ and, in particular, quadratic gravity. In such models, e.g., \cite{qua1, qua2, qua3},
  it is supposed that quadratic corrections to GR become important at energy scales of 
  Grand Unification theories, which approximately corresponds to  $a_2 \sim 10^{10} r_0^2$ in 
  \rf{f-R}, where $r_0$ is of the order of the Planck length. The same assumption was used
  in some models predicting a semiclassical bounce instead of a Schwarzschild singularity
  inside a black hole \cite{bou1, bou2}. Now, assuming that the curvature $R$ is of the order
  $R \sim \e^{-2\beta}/r_0^2 = x^2/r_0^2$ and substituting this into the condition \rf{Rsmall},
  we obtain 
\beq       \label{x5}
		x^2 \ll r_0^2/a_2 =  10^{-10} \qquad \Longrightarrow \qquad x \ll 10^{-5}.  		
\eeq 
  
  With such values of $x$ and the above-mentioned values of $k_C$, it is clear that the 
  Casimir contribution to the potential \rf{W-x} is also quite negligible, and we can restrict 
  $W(x)$ to the first two terms.
  
  Other restrictions on admissible values of $x$ follow from two evident conditions:
  (i) A classical space-time description requires that the size of the extra dimensions,
  $r = r_0 \e^\beta = r_0/x$  should be much larger than the fundamental length 
  $r_0 = 1/\mD$, hence $x \ll 1$; (ii) This size should be small enough for the extra
  dimensions to be invisible for modern instruments, that is, 
  $r = r_0/x \lesssim 10^{-17}$ cm, approximately corresponding to the TeV energy scale.
  If we assume that $\mD \sim m_4 \sim 10^{-5}\rm\ g \sim 10^{33} \ cm^{-1}$, it follows 
  $x \gtrsim 10^{-16}$. Admitting that $\mD$ may be some orders of magnitude smaller
  than $m_4$ and recalling \rf{x5}, we can more or less safely assume
\beq  			\label{x-ord}
 			10^{-13} \lesssim x \ll 10^{-5}.
\eeq  
  
  For a further study and further estimates, it is necessary to make an assumption on a 
  conformal frame in which we interpret the observations. This choice ultimately depends on 
  how fermions are included in a (so far unknown) underlying unification theory of all physical
  interactions. We will consider two options: the Jordan frame with the action \rf{act2}, 
  directly derived from the original D-dimensional theory, and the Einstein frame
  with the action \rf{act3}.
  
\subsection*{The Einstein frame}  

  Assuming that our observed space-time corresponds to the Einstein frame, we have 
  \eqs \rf{E00} and \rf{E-tr} for the scale factors $a(t)$ and $\e^{\beta(t)}$. The Hubble 
  parameter of the present Universe, $H(t) = {\dot a}/a$, is of the order 
  $10^{-17}\,{\rm s}^{-1} \approx 10^{-61} r_0^{-2}$ if $r_0 = 1/m_4$; evidently,
  $\dot\beta$ should be at most of the same order, therefore the same is required from 
  $V(\beta)$. Then it follows from \eqn{E00}
\beq               \label{Eest}
  			\lambda x^n - \Half n(n-1) x^{n+2} \lesssim 10^{-122}.
\eeq  
  Assuming $x = \e^{-\beta} \sim 10^{-10}$ (well in the range \rf{x-ord}), 
  and using \eqn{Eest}, we can consider the following two options:
\medskip  
  
  (i) $\lambda = 0, \ \ \ n(n-1) \cdot 10^{-10 n-2} \lesssim 10^{-122}\ \ \then \ \ n \geq 12$.
  \medskip
  
  (ii) $\lambda \ne 0$, then we can rewrite \rf{Eest} as 
\[  
  		\lambda = \Half n(n-1)\cdot 10^{-20} + O(10^{-122 + 10 n}),
\]  
  thus the estimate of  $\lambda$ depends on $n$ which can now be smaller than 12, 
  but a considerable fine tuning is necessary, for example, 
\bearr
		n = 10 \ \ \then \ \ 	\lambda = 45 \times 10^{-20} \pm O(10^{-22}),
\nnnv  
 		\ n = 5 \ \ \then \ \ 	\lambda = 10^{-19} \pm O(10^{-72}).
\earn  
  In all cases, \eqs \rf{E00} and \rf{E-tr} should be solved numerically.
 
\subsection*{The Jordan frame}  

  If we assume that the observed space-time metric is $ds_J^2 = \e^{-n\beta} ds_E^2$,
  we can write down the cosmological metric as
\beq                         \label{ds_J}
		ds_J^2 = \e^{-n\beta} \Big[dt^2 - a^2(t) d{\vec x}^2\Big] = d\tau^2 - a_J^2(\tau) d{\vec x}^2,
\eeq  
  where $\tau$ is cosmological time such that $d\tau = \e^{-n\beta/2} dt$, and 
  $a_J(\tau) = \e^{-n\beta/2} a(t)$ is the Jordan-frame scale factor.
  
  Now, the Hubble parameter is defined as
\beq       \label{H_J}
		H = \frac 1{a_J}\, \frac{d a_J}{d\tau} 
			= \bigg(\frac{\dot a}{a} - \Half n \dot\beta\bigg)\e^{-n\beta/2},
\eeq   
  which leads to the estimate 
\beq               \label{J1}
		\frac{\dot a}{a} - \Half n \dot\beta \sim H x^{n/2}. 
\eeq
  On the other hand, $\dot\beta$ is subject to the observational constraint on possible variations
  of the effective gravitational constant $G$ (it is a true constant in the Einstein frame but 	
  varies in Jordan's proportionally to $\e^{-n\beta}$, see \rf{act2}): according to 
  \cite{var-G1, var-G2}, we must have $(1/G)|dG/d\tau| \lesssim 10^{-3} H$, whence it follows
\beq                  \label{J2}
			|\dot \beta| \lesssim 10^{-3} H x^{n/2}.
\eeq 
   It means that we can neglect the second term in \rf{J1} and simply take $\dot a/a \sim H x^{n/2}$.
   Applying this to \eqn {E00} in the same manner as in the Einstein frame, we arrive at the 
   relation
\beq                  \label{J3}
			\lambda - \Half n(n-1) x^2 \sim 10^{-122},
\eeq   
   which means, for any reasonable choice of $x$, an unnatural fine tuning of the value of 
   $\lambda$. Recalling that $x = \e^{-\beta}$ varies with time while $\lambda = \const$,
   we come to the conclusion that the Jordan frame does not lead to a plausible cosmology 
   in the present statement of the problem. 
    
\section{The metric of a region with $R\to 0$. Numerical simulations}

  The main result of the previous discussion is the following: if an observer is inside such a region, 
  then an unnaturally strong fine tuning of the model parameter $\lambda$ is necessary
  (except for $n \geq 12$ in the Einstein frame). Also, the 
  first and second-order derivatives of the potential are not defined at the point $\phi\equiv R_n\to 0$,
  so that there are no oscillations around such a minimum. But such oscillations are necessary for 
  a successful reheating just after inflation. Therefore, our Universe is hardly described by the metric 
  discussed above.
  
  Nevertheless, such regions could exist somewhere in the Universe, probably not too close to our 
  Milky Way galaxy, otherwise it could disturb too strongly the well-known observable picture. 
  It is a separate problem how we, being external observers for such regions, could detect 
  signals coming from there.  Let us note that a similar problem was discussed some years 
  ago concerning possible large-scale antimatter regions \cite{KhlopovRubin,Khlopov:2000as}.
  
  It looks worthwhile to analyze the possible metric inside such regions.
  The first field equation for $\beta(t)$ (excluding $a(t)$ with \rf{E00}) reads
\beq                               \label{scalareq}
	\ddot\beta + \dot\beta\sqrt{3\left[\frac12 n(n+2)\dot\beta^2+V(\beta)\right]}
			 +\frac 1{ n(n+2)}V_{\beta}=0 	
\eeq
  The second equation \rf{E00} is necessary for obtaining the 4D scale factor. 
  
  We are not restricted by the observational arguments if a region in question is much smaller than 
  the whole Universe, but it should be assumed to be large and homogeneous enough, so that the 
  cosmological metric \rf{ds_E} could be applicable. On the other hand, no such severe fine tuning 
  is needed there for the model parameters since it is not necessary to require an extreme 
  smallness of the extra dimensions.
  
  The field motion near a nonzero minimum of $V(\phi)$ in Fig.\,1 corresponds to the 
  observable inflation for specific parameter values  \cite{Fabris:2019ecx}. 
  We need only one of them, $\lambda =\Lambda_D = 0.0125$. 
  Also, we put $V(\beta ) = \Lambda_D e^{-n\beta}$, see \eqref{V-beta}. A numerical solution
  to these equations is presented in Fig.\,2. One can see that the extra dimensions expand 
  very slowly --- we show the dynamics during the most interesting time interval from the 
  sub-Planckian time scale to the post-inflationary period ($10^9 \sim 10^{-34}$ s in our units). 
  The behavior of the curves remains roughly the same up to the present time. It is of interest 
  that the expansion rate strongly depends on the number $n$ of extra dimensions: it is 
  inversely proportional to $n$. 
\begin{figure}
\centering
\includegraphics[width=0.55\linewidth]{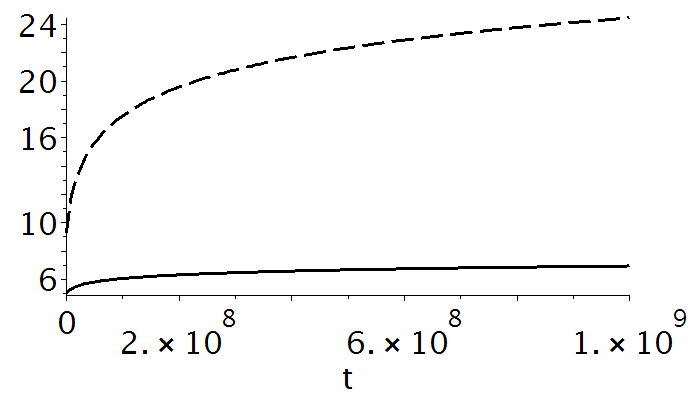}
\caption{\small
	Time dependence of the size of extra space (solid) and the main 4D space (dashed) in a
	logarithmic scale; $n=5, \Lambda_D=0.0125$.}
	\label{fig:evolution}
\end{figure}

  It was the Einstein frame. A transition to the Jordan frame, written explicitly using \eq \eqref{ds_J}, 
  gives no new results simply because of a very slow variation of the extra-space metric.

\section{Conclusion}

  The space domains containing the expanding extra dimensions like those described above 
  could exists in our Universe.
  It would be of interest to study light propagation from there. Also, ordinary matter could penetrate 
  inside a domain possessing unusual properties, or they could be connected with our space 
  through wormholes \cite{ex-wh1, ex-wh2, 2016PhLB..759..622R}  
  It is also of interest to study the stability of such regions from the point of view of an external 
  observer: do they expand or shrink? 

  Another possibility arises if we recall that the potential maximum separates two 
  minima as in Fig.\,1. In this case, a closed wall is formed \cite{Khlopov:1998nm}, which 
  could expand or shrink, sweeping out the internal domain. 
  As was discussed in \cite{BR-book,Belotsky:2018wph,2000hep.ph....5271R}, shrinking walls could 
  cause multiple formation of black holes in the Universe. It could solve the problem of primordial 
  black hole formation \cite{KhlopovRubin,Khlopov_2010}.
  
\subsection*{Acknowledgment}
  	
  This work has been supported by the Ministry of Science and 
  Higher Education of the Russian Federation, Project ``Fundamental properties of 
  elementary particles and cosmology'' N 0723-2020-0041.
  The work of S.R/ was also supported by RFBR grant No. 19-02-00930,
  and the work of K.B. by RFBR grant No. 19-02-00346.


\end{document}